# TIME VARIATIONS OF THE SOLAR NEUTRINO FLUX DATA FROM SAGE AND GALLEX-GNO DETECTORS BY SIMPLE DENOISING ALGORITHM USING WAVELET TRANSFORM

**Koushik Ghosh[1] and Probhas Raychaudhuri[2]**

**[1]Department of Mathematics**
**Dr. B.C. Roy Engineering College**
**Jemua Road, Fuljhore**
**Durgapur-713 206**
**INDIA**
**E-mail:koushikg123@yahoo.co.uk**

**[2]Department of Applied Mathematics**
**University of Calcutta**
**92, A.P.C. Road, Calcutta-700 009**
**INDIA**
**E-mail:probhasprc@rediffmail.com**

## ABSTRACT:

We have used Simple Denoising Algorithm using Wavelet Transform on the monthly solar neutrino flux data from (1) SAGE detector during the period from 1st January 1990 to 31st December 2000; (2) SAGE detector during the period from April 1998 to December 2001; (3) GALLEX detector during the period from May 1991 to January 1997; (4) GNO detector during the period from May 1998 to December 2001; (5) GALLEX-GNO detector (combined data) from May 1991 to December 2001 and (6) average of the data from GNO and SAGE detectors during the period from May 1998 to December 2001. (1) exhibits periodicity around 20, 27, 36, 70, 90 and 114 months. (2) shows periodicity around 21, 28, 31 and 33 months. For (3) we observe periodicity around 24, 40, 54, 57, 59, 62 and 64 months. For (4) periodicity is seen around 23, 30 and 34 months. (5) gives periodicity around 49, 57, 64, 74, 79, 88, 95 and 99 months while (6) shows periodicity around 24, 32, 35 and 38 months. We have found almost similar periods in the solar flares, sunspot data, solar proton data ( $\overline{\in}$>10 Mev) which indicates that the solar activity cycle may be due to the variable character of nuclear energy generation inside the sun.

## I. INTRODUCTION:

Solar neutrino flux detection is very important towards the understanding of solar internal structure (i.e. its nuclear energy generation process, temperature, density etc.). Standard solar model (SSM) does not predict solar neutrino flux variation [1]. Only perturbed solar model of Raychaudhuri [2] suggested that the solar neutrino flux should vary with the solar activity cycle. Solar neutrino flux data from Homestake detector varies with the solar activity cycle [3, 4, 5, 6] but at present it appears that there is no significant anticorrelation of solar neutrino flux data with the sunspot numbers. Raychaudhuri [7, 8] showed that solar neutrino flux data from Kamiokande, Superkamiokande, GALLEX and SAGE are varying with the solar activity cycle. From the analysis it appears that variation of solar neutrino flux data indicates that there must be periodicity in the data.

Many authors analysed the solar neutrino flux data from Homestake detector and have found short-time periodicities around 5, 10, 15, 20, 25 months etc. [9, 10, 11, 12]. Raychaudhuri analysed the solar neutrino flux data from SAGE, GALLEX and Superkamiokande and have found that solar neutrino flux data varies with the solar activity cycle and have found periodicities around 5 and 10 months.

Applications of wavelets and multiresolution analysis to reaction engineering systems from the point of view of process monitoring, fault detection, system analysis and so on, is an important topic and of current research interest [13]. The presence of noise in a time-series data restricts one significantly to have meaningful information from the data. Noise in experimental data can also cause misleading conclusions [14]. Donoho and Johnstone [15] introduced two types of Denoising: linear Denoising and nonlinear Denoising. In linear Denoising noise is assumed to be concentrated only on fine scales and that all the wavelet coefficients below these scales are cut off. Nonlinear Denoising, on the other hand, treats noise reduction by either cutting off all coefficients below a certain threshold or reducing all coefficients by this threshold [13, 15]. The threshold values are obtained by statistical calculations and have been seen to depend on the standard deviation of the noise [16].

Roy et al [13] proposed a denoising algorithm making use of the discrete analogue of the Wavelet Transform (WT), which in many ways complements the well-known Fourier Transform (FT) procedure. The unique part of this proposal is that in this noise-reduction algorithm the threshold level for noise is identified automatically. We here extend this useful and advantageous algorithm to search for the time variation in solar neutrino flux data obtained from different detectors. We have used this Simple Denoising Algorithm on the monthly solar neutrino flux data from (1) SAGE detector during the period from 1st January 1990 to 31st December 2000; (2) SAGE detector during the period from April 1998 to December 2001; (3) GALLEX detector during the period from May 1991 to January 1997; (4) GNO detector during the period from May 1998 to December 2001; (5) GALLEX-GNO detector (combined data) from May 1991 to December 2001 and (6) average of the data from GNO and SAGE detectors during the period from May 1998 to December 2001. For this method graphs are plotted to depict the periodic nature of the solar neutrino flux data and the corresponding periodicities are calculated.

## THEORY: THE NOISE REDUCTION ALGORITHM BASED ON DISCRETE WAVELET TRANSFORM:

The noise-reduction algorithm consists of the following steps [13]:

**Step 1:** In the first step, we differentiate the noisy signal x(t) to obtain the data $x_d(t)$, using the central finite difference method with fourth order correction to minimize the error [13, 17] i.e.

$$x_d(t)=dx(t)/dt \tag{1}$$

**Step 2:** We then take the Discrete Wavelet Transform of the data $x_d(t)$ and obtain wavelet coefficients $W_{j,k}$ at various dyadic scales j and displacements k. A dyadic scale is the scale whose numerical magnitude is equal to 2 rose to an integer exponent and is labelled by the exponent. Thus the dyadic scale j refers to a scale of size $2^j$. In other words, it indicates a resolution of $2^j$ data points. Thus a low value of j implies a finer resolution, while high j analyzes the signal at a larger resolution. This transform is the discrete analogue of continuous Wavelet Transform [13, 18] and it can be represented by the following formula

$$W_{j,k} = \int_{-\infty}^{\infty} x_d(t)\ \Psi_{j,k}(t)\, dt \tag{2}$$

with

$$\Psi_{j,k}(t)= 2^{j/2}\Psi(2^j t-k)$$

Where j, k are integers. As for the wavelet function $\Psi(t)$, we have chosen Daubechies' compactly supported orthogonal function with four filter coefficients [19, 20]. Here for simplicity we take

$$\Psi_{j,k}(t)= 2^{j/2} (2^j t-k)^2 \quad (3)$$

**Step 3:** In this step we estimate the power $P_j$ contained in different dyadic scales j, via

$$P_j(x)= \sum_{K=-\infty}^{\infty} |W_{j,k}|^2 \qquad (j=1, 2, \ldots.) \quad (4)$$

By plotting the variation of $P_j$ with j, we see that it is possible to identify a scale $j_m$ at which the power due to noise falls of rapidly. This is important because, as we shall see from the studies on the case examples, it provides an automated means of detecting the threshold. Identification of the scale $j_m$ at which the power due to noise shows the first minimum allows us to reset all $W_{j,k}$ up to scale index $j_m$ to zero, that is $W_{j,k}=0$, for j=1, 2,…., $j_m$ [13].

**Step 4:** In the fourth step, we reconstruct the denoised data $\hat{x}_d(t)$ by taking the inverse transform of the coefficients $W_{j,k}$

$$\hat{x}_d(t)= \sum_{j=0}^{\infty} \sum_{k=-\infty}^{\infty} \hat{W}_{j,k} \Psi_{j,k}(t) \quad (5)$$

This set of obtained $\hat{x}_d(t)$ gives measure of time variation in the signal. Upon differentiation the contribution due to white noise moves towards the finer scales because the process of differentiation converts the uncorrelated stochastic process to a first order moving average process and thereby distributes more energy to finer scales [13]. Finally we plot the graph of $\hat{x}_d(t)$ vs. t to obtain the corresponding peaks.

## RESULT:

For numerical simulation we take j=1, 2, …., 5 and k=1, 2, …., 5. Calculations yield different periodicities for the six types of solar neutrino flux data in the present analysis. An interesting observation for the presently considered algorithm is that the method has a tendency to locate the periods in comparatively larger time scale. So we search for the initial peaks of the solar neutrino flux data by taking the initial time range as a whole and then restudy the case to obtain the initial peaks. The peaks obtained for the presently considered six data are presented in the tabular form:

| **Experiments** | **Periodicities (in months)** |
|---|---|
| (1) SAGE detector during the period from 1st January 1990 to 31st December 2000 | 20, 27, 36, 70, 90 and 114 |
| (2) SAGE detector during the period from April 1998 to December 2001 | 21, 28, 31 and 33 |
| (3) GALLEX detector during the period from May 1991 to January 1997 | 24, 40, 54, 57, 59, 62 and 64 |
| (4) GNO detector during the period from May 1998 to December 2001 | 23, 30 and 34 |
| (5) GALLEX-GNO detector (combined data) from May 1991 to December 2001 | 49, 57, 64, 74, 79, 88, 95 and 99 |
| (6) average of the data from GNO and SAGE detectors during the period from May 1998 to December 2001 | 24, 32, 35 and 38 |

**FIGURE 1: ANALYSIS OF SAGE DATA BY SIMPLE DENOISING ALGORITHM USING WAVELET TRANSFORM**

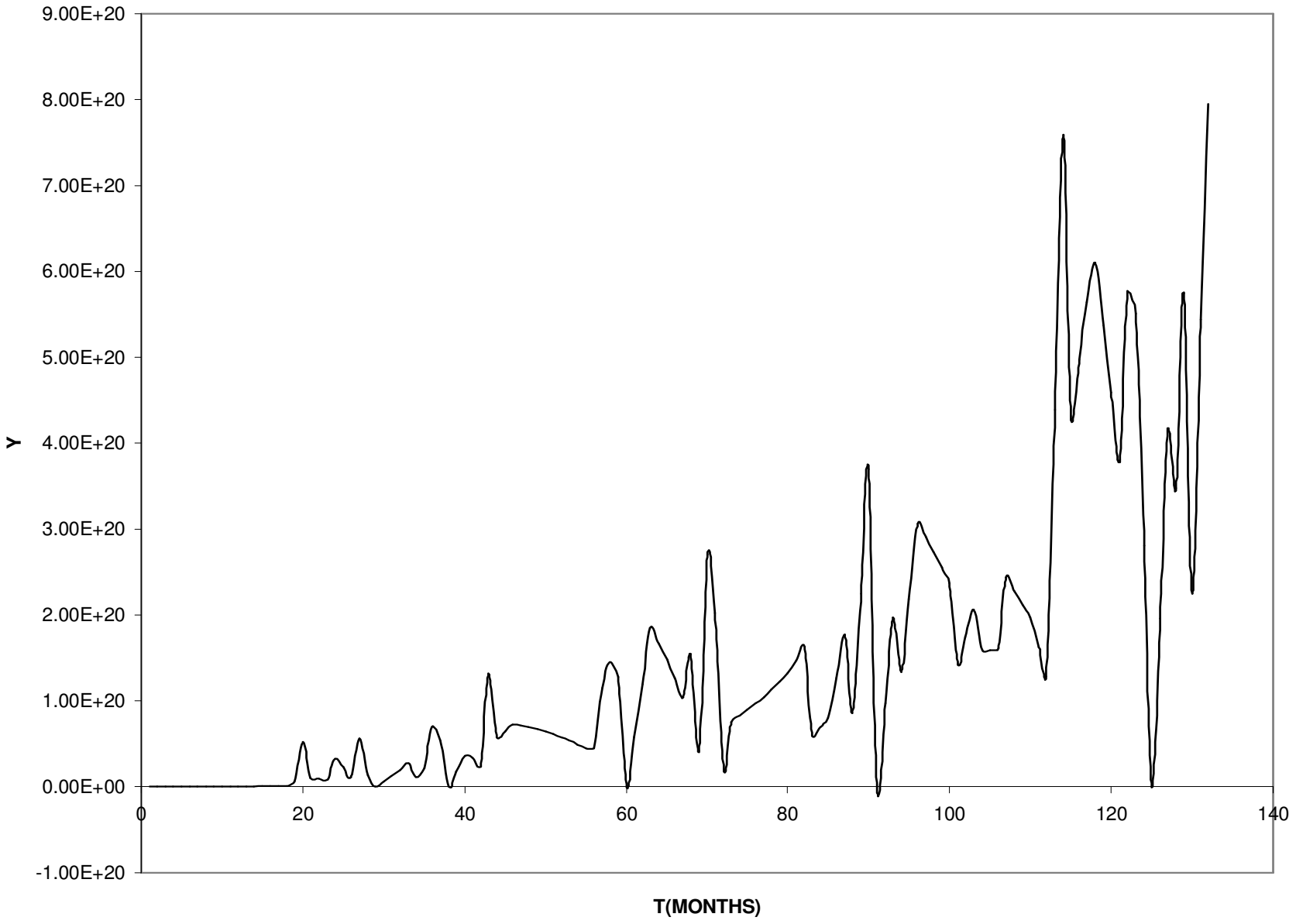

**FIGURE 2: ANALYSIS OF SAGE DATA BY SIMPLE DENOISING ALGORITHM USING WAVELET TRANSFORM**

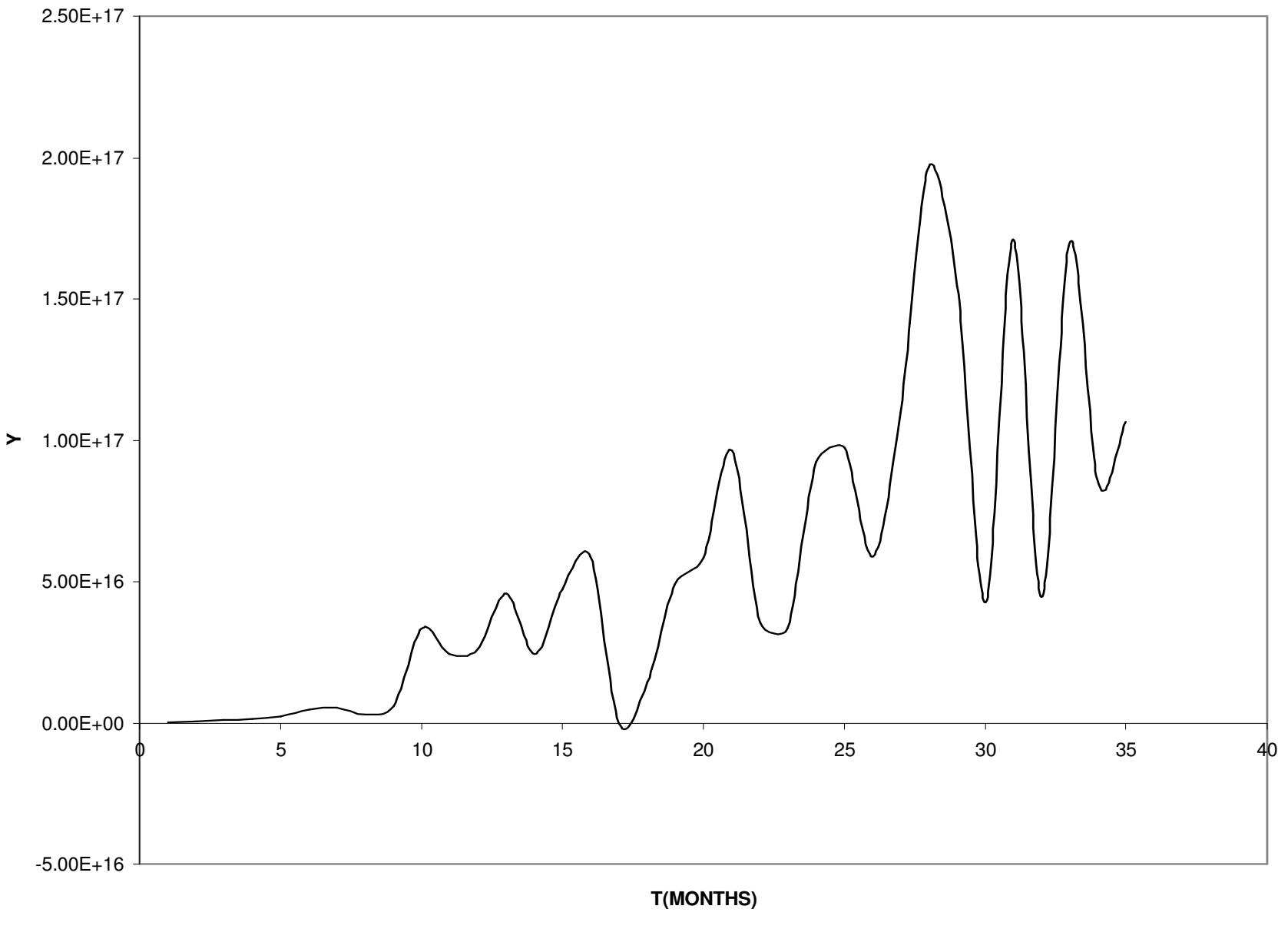

FIGURE 3: ANALYSIS OF GALLEX DATA BY SIMPLE DENOISING ALGORITHM BY USING WAVELET TRANSFORM

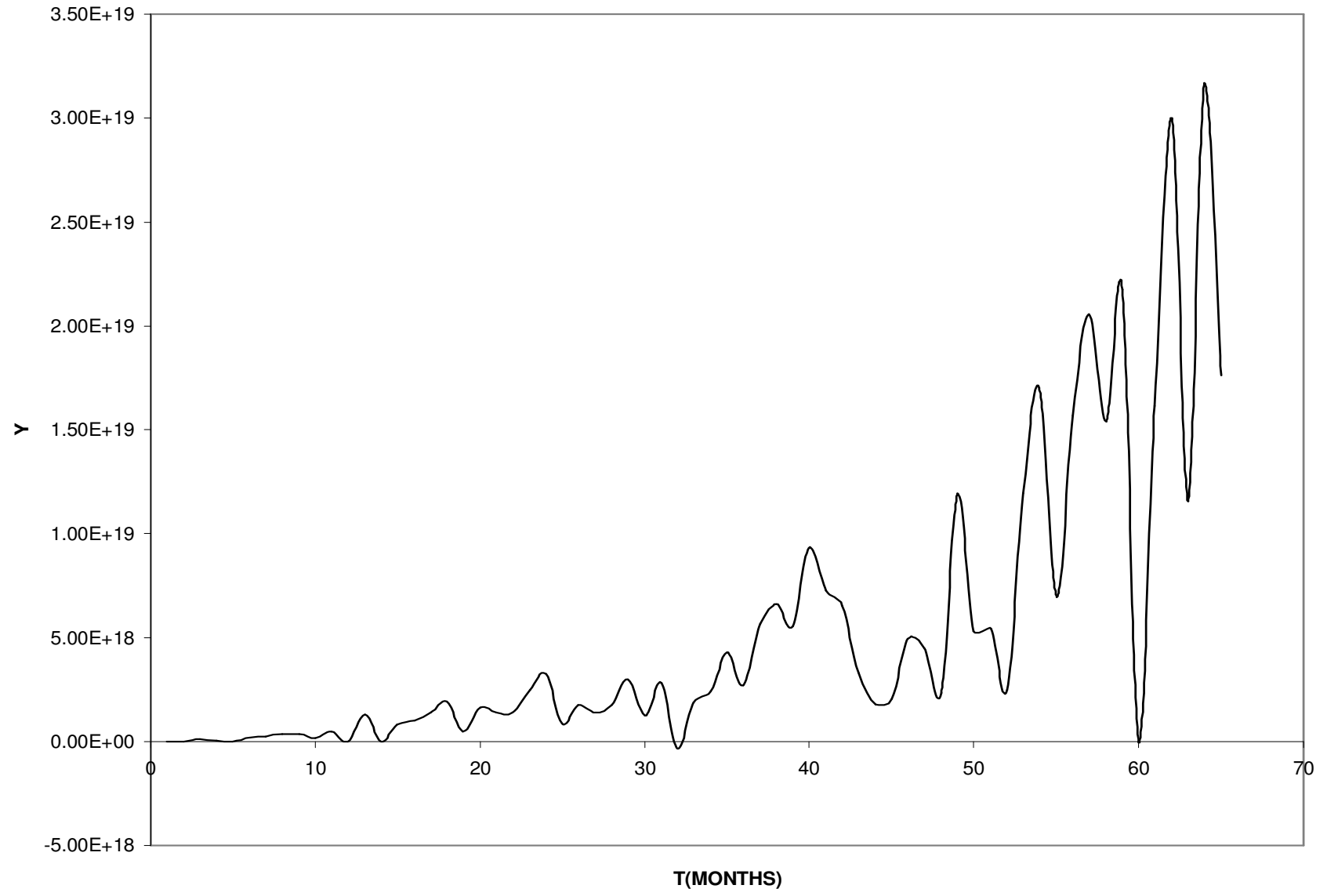

FIGURE 4: ANALYSIS OF GNO DATA BY SIMPLE DENOISING ALGORITHM USING WAVELET TRANSFORM

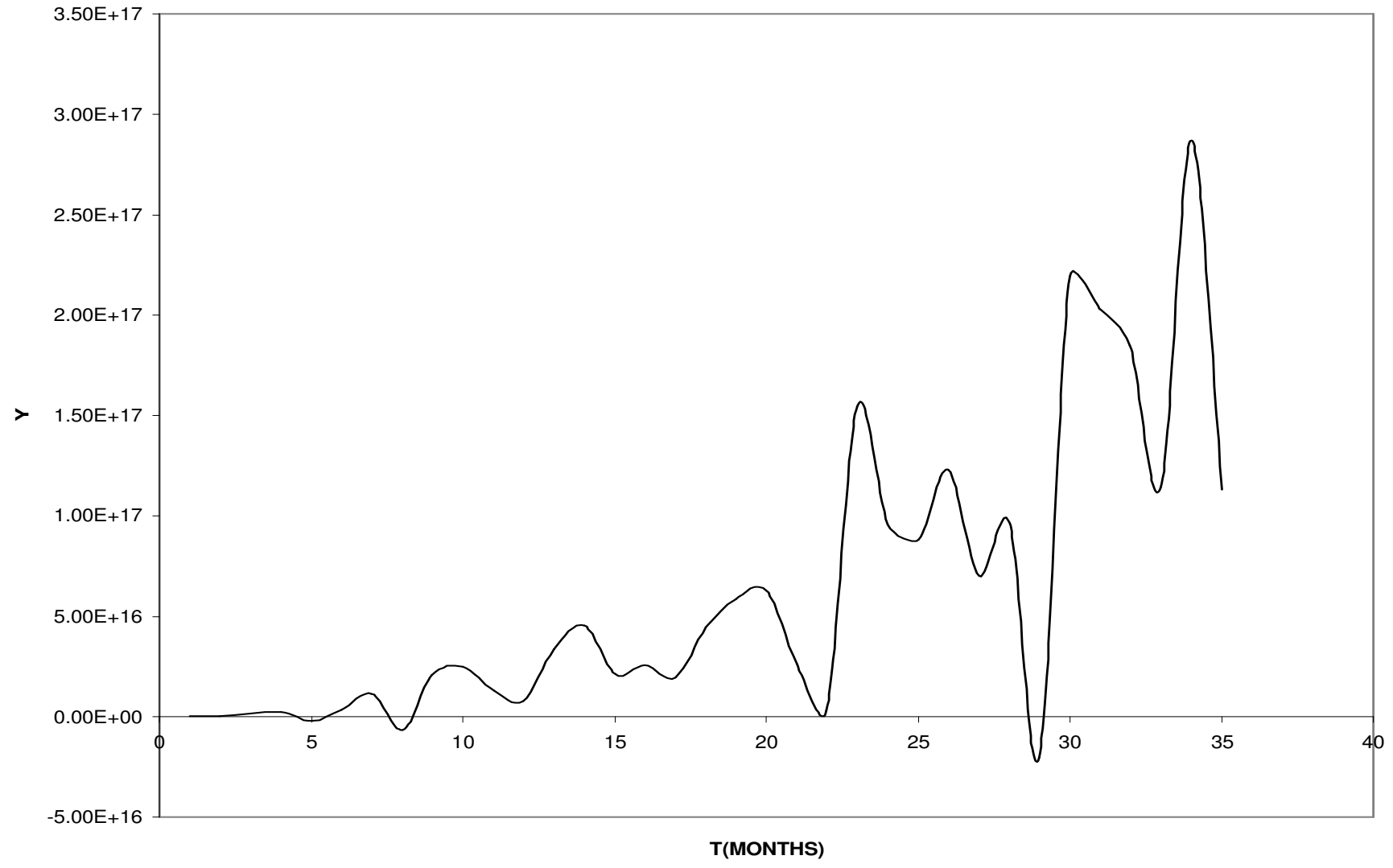

FIGURE 5: ANALYSIS OF COMBINED GALLEX-GNO DATA BY SIMPLE DENOISING ALGORITHM USING WAVELET TRANSFORM

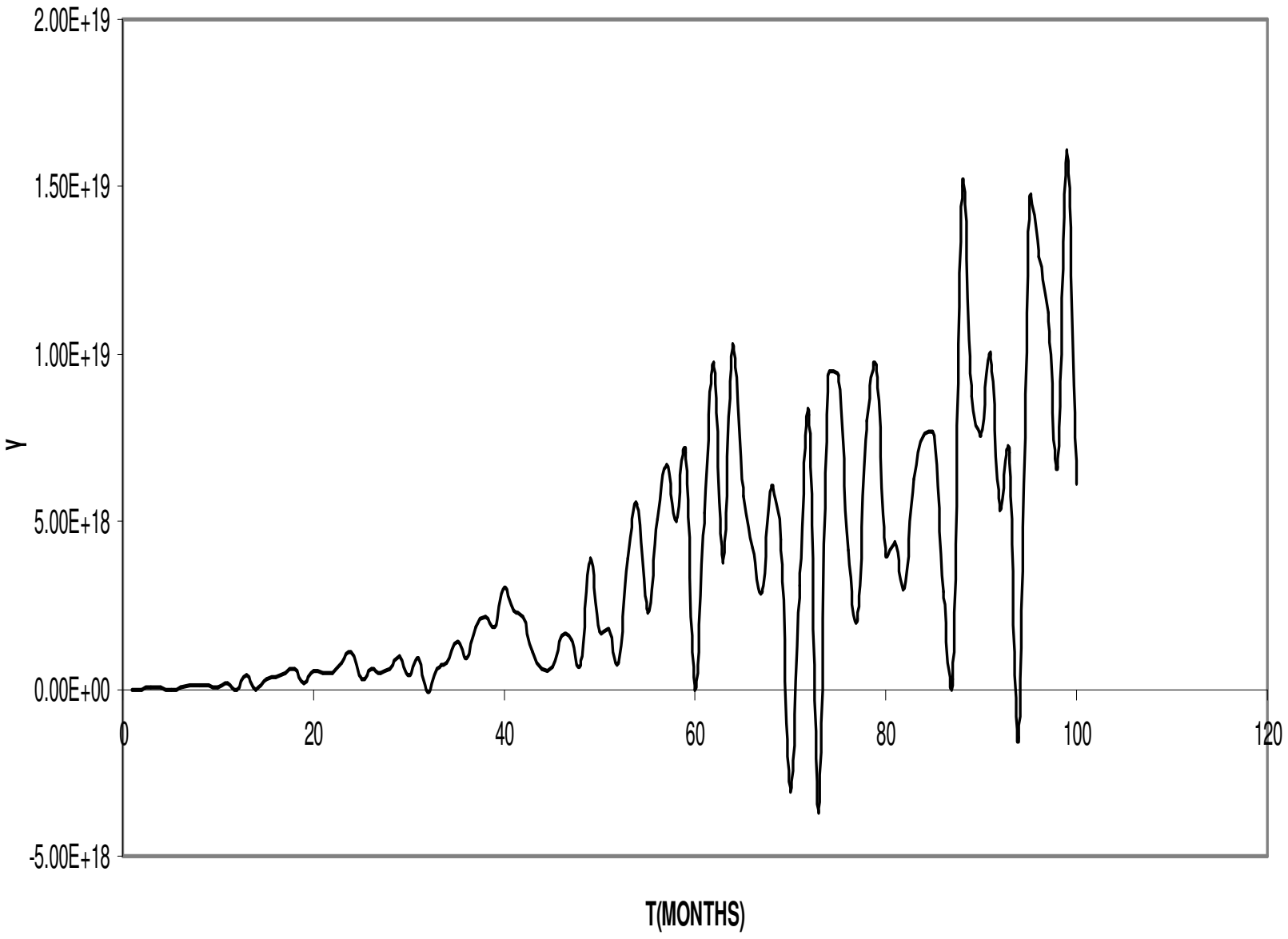

FIGURE 6: ANALYSIS OF AVERAGE OF GNO AND SAGE DATA BY SIMPLE DENOISING ALGORITHM BY USING WAVELET TRANSFORM

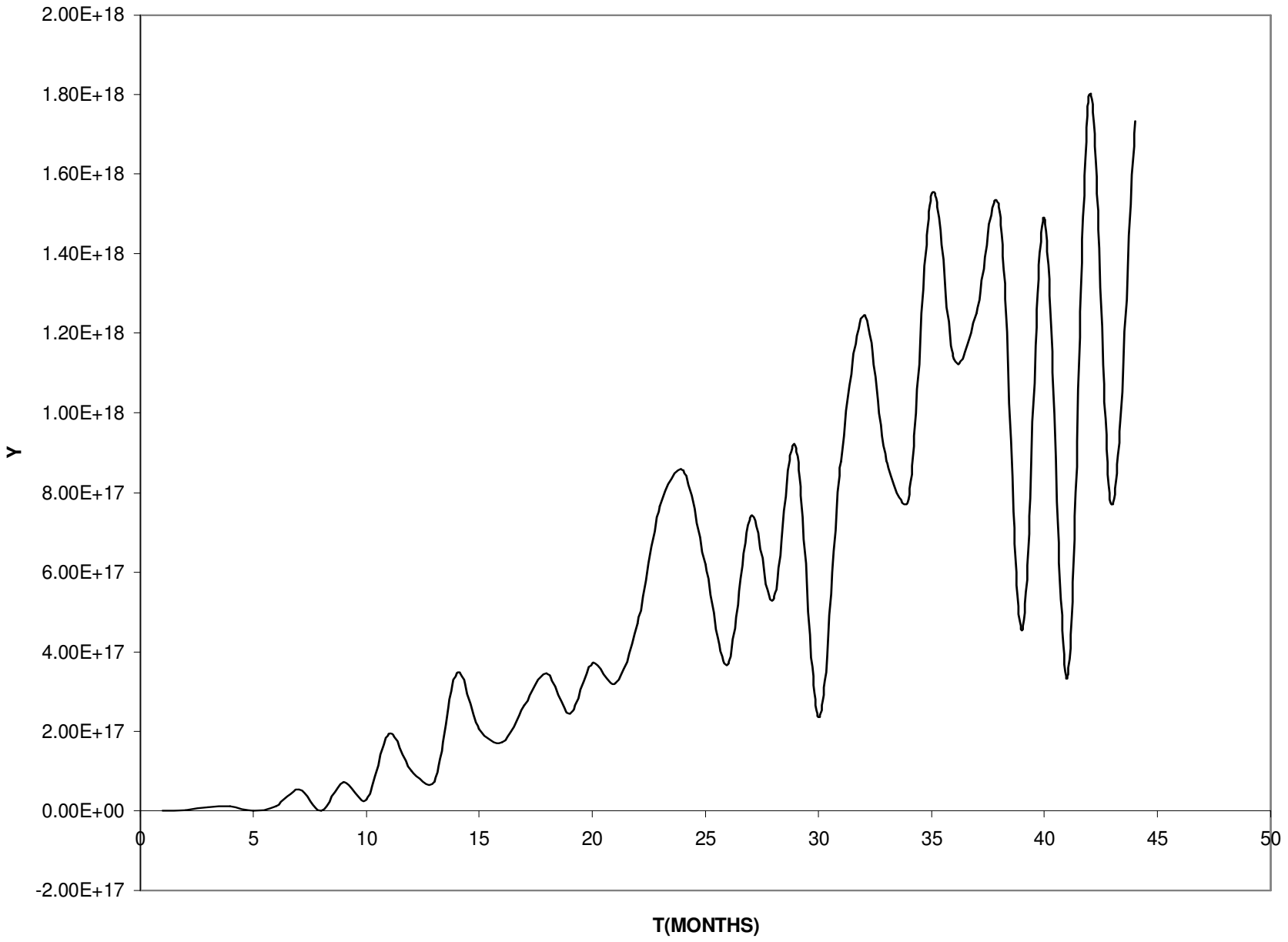

**CONCLUSION:**

The obtained period of 20 months for the data (1) is not appreciably different from the period of 19 months obtained for the same data by Ferraz-Mello Method of Date Compensated Discrete Fourier Transform in our one of earlier communications [21]. We have also obtained a period of 27 months for data (1) by the Ferraz-Mello Method [21] which is significantly comparable with the obtained period of 24 months for data (1) by the present analysis. In our analysis of the solar neutrino flux data by Rayleigh Power Spectrum [22] we obtained periods of 18.7 and 32.9 months for data (1) which are comparable with the obtained peaks of 20 and 36 months for the same data by the present analysis. For data (3) the presently obtained period of 24 months is almost similar to the peak of 26.9 months obtained for the same data by Rayleigh Power Spectrum Analysis [22].